\documentstyle[11pt,newpasp,twoside]{article}
\def\edcomment#1{\iffalse\marginpar{\raggedright\sl#1\/}\else\relax\fi}
\reversemarginpar

\begin{document}
\title{The Assembly and Evolution of Spiral Disks}

% \vskip -0.1in

\author{Matthew Bershady$^1$ and David Andersen$^2$}
\affil{$^1$Department of Astronomy, U. of Wisconsin--Madison, USA\\
$^2$Dept. of Astronomy \& Astrophysics, Penn State University, USA}

\begin{abstract}

We explore how the growth rate of spiral disks can be measured via
analyses of the scatter in the Tully-Fisher (TF) relation of local and
intermediate redshift galaxies. As an initial step, we show it is
possible to construct a low-dispersion TF relation for nearly face-on,
nearby spirals. We find these spiral disks are non-circular, (a mean
ellipticity of 6\%), which accounts for $\sim$0.1 mag of the intrinsic
scatter in the nearby TF relation of even the most ``normal'' looking
spirals. If this ellipticity is induced by matter accretion, we expect
to find greater disk ellipticity or disturbance in the past. We find
evidence that more extreme outliers of the intermediate-redshift TF
relation are more morphologically and kinematically disturbed. Whether
this effect reflects a redshift trend or selection bias of local
samples needs to be addressed.

\vskip -0.25in
\end{abstract}

\section{Measuring the Growth of Large Disks}

\vskip -0.1in

Dynamically cold, large galaxy disks seen today appear to have formed
early or gradually; they are easily destroyed through strong
interactions or mergers. For example, Wyse (these proceedings) argues
the Milky Way has not suffered a major merger in the last 10 Gyr. The
nearly constant co-moving number of large disks out to $z \sim 1$
(e.g. Lilly et al. 1998) and the subtle changes with redshift in the
scaling relations for large, field spirals corroborate a picture of
early formation or slow, quiescent growth, but infrequent death.

The mass accretion rate onto disks is not known directly, however, at
any epoch. Direct measurements of disk mass as a function of time,
e.g., from measurement of disk scale-heights and stellar velocity
dispersions, are desirable, but currently unavailable -- except for a
handful of local systems. Photometric estimates of disk mass based on
colors are uncertain and incomplete. Another approach is to use the
evolution of the zeropoint and scatter in disk scaling-relations as a
diagnostic of {\it changes} in mass accretion rates. Here we focus on
the correlations with, and cause of the scatter in the Tully-Fisher
relation.

For gradual formation, we assume that discrete, but small accretion
events perturb otherwise axisymmetric systems. Were we to obtain a
complete census of galaxies at difference epochs, the relative number
of perturbed systems should then reflect changes in the minor-merger
rate. Zaritsky \& Rix (1997) estimated the minor-merger rate for
nearby galaxies by comparing the amplitude of photometric $m=1$ modes
to $B-R$ and offsets in luminosity from a fiducial TF relation.
Kinematic measures of disturbance would be preferred, however, since
the long-term morphological perturbation may be subtle (Haynes 2000),
and photometric effects may evolve with look back-time due to changing
gas fractions and star-formation rates (Mihos, these
proceedings). Here we give two examples of how spatially resolved
kinematics can directly link disk asymmetries to scatter in the TF
relation. This, in turn, points to a tractable method for measuring
the growth rate of spiral disks.

% Whether changes in disk mass accretion rates via signatures of increased
% asymmetry, and scatter, 

% correlate with color and asymmetries and in the disk light distributions

\begin{figure}
\plotfiddle{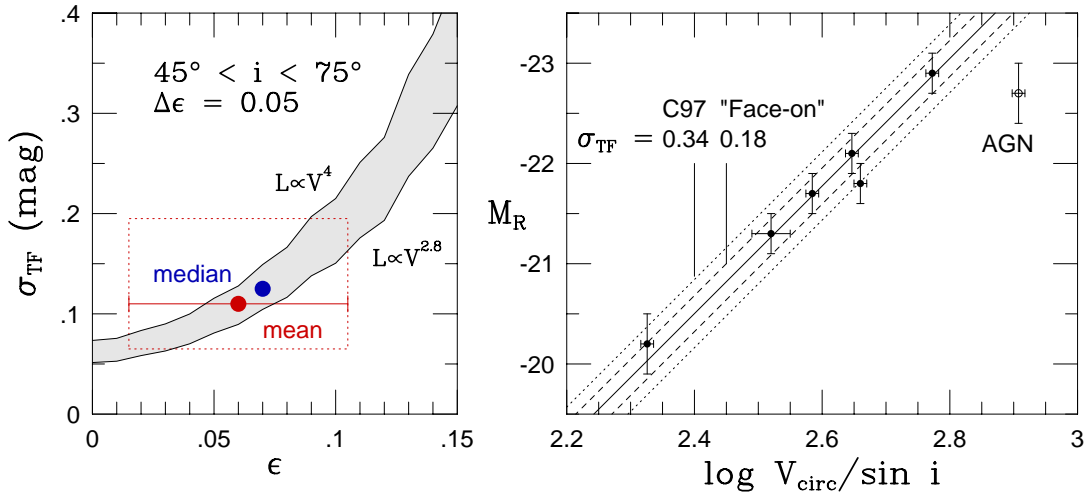}{2in}{0}{55}{55}{-210}{-75}
\vskip 0.15in
\caption{\hsize 5.2in \baselineskip 0.165in (a) The contribution to
the TF scatter ($\sigma_{TF}$) due to inclination errors produced by
intrinsic disk ellipticity ($\epsilon$), for two TF slopes (labeled).
The mean, standard deviation and median ellipticity of our sample are
indicated. (b) The $R$-band TF-relation for our nearly face-on sample,
compared to Courteau's (1997) zeropoint and scatter (assuming
$r-R=0.36$). The one outlier has an AGN and W$_{50}$/W$_{20}$
line-width ratios consistent with a dynamically disturbed system
(Conselice et al. 2000).}
\vskip -0.15in
\end{figure}

\vskip -0.35in
\null

\section{Non-Circular Disks and the Local Tully-Fisher Error Budget}

\vskip -0.1in

Using high signal-to-noise, integral-field, echelle spectroscopy and
surface photometry of seven apparently face-on spirals, we have been
able to construct H$\alpha$ velocity fields out to 3 scale lengths,
and compare kinematic to photometric position angles, and kinematic
inclinations to photometric axis-ratios (Andersen et al. 2000, and
these proceedings). From mismatches between these quantities, we find
that normal, non-barred, intermediate-type spiral disks are
non-circular, with a model-dependent estimate of mean ellipticity at
the 6\% level. Considering the effects of our observed distribution of
intrinsic ellipticity only on the inferred photometric inclination, we
estimate this accounts for $\sim0.1$ mag of scatter in the TF-relation
for samples selected within $45^\circ<i<75^\circ$ (Figure 1a),
consistent with Franx \& de Zeeuw's expectations (1992).

Evidence that intrinsic ellipticity predominantly effects photometric
but not kinematic measures of inclination is shown in Figure 1b. We
establish a tight, $R$-band TF relation for galaxies with
$16^\circ<i<32^\circ$ by using kinematic inclinations. Not only is our
zeropoint in agreement with Courteau (1997), but our scatter is
smaller (albeit with a much smaller sample). Hence, it is now
possible to study the TF relation for nearly face-on samples, where
photometric projection effects and internal extinction are minimized,
and the perpendicular component of the disk velocity dispersion can be
measured to estimate disk mass.

\vskip -0.3in
\null

\section{Scatter in the Tully-Fisher Relation at Intermediate Redshifts}

\vskip -0.1in

An outstanding question is whether local disk non-circularity is caused
primarily by small amounts of lopsided matter accretion, triaxial
halos, or other dynamical processes.  In principle one might hope to
determine this from a detailed analysis of disk velocity fields.
Alternatively, we look here to higher-redshift systems to see if
greater ellipticity or asymmetry is evident, as we might expect if
minor mergers are the dominant source of this perturbation.

% define this here:

\begin{figure}
\plotfiddle{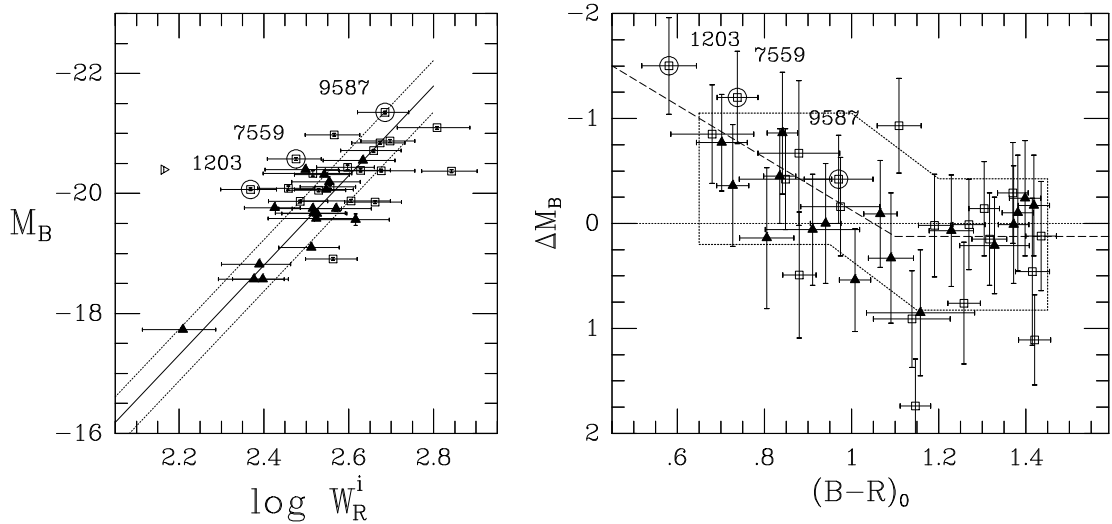}{2in}{-90}{60}{60}{-220}{245}
\vskip 0.35in
\caption{\hsize 5.2in \baselineskip 0.165in (a) $B$-band Tully-Fisher
(TF) Relation for galaxies between $0.05<z<0.4$ (H$_0 = 68$, $\Omega_0 =
1$) compared to a local calibration (lines; Pierce \& Tully 1992). (b)
Residuals of this TF relation versus rest-frame $B-R$ color. The
enclosed area indicates the relation found for local galaxies. For
both panels: filled triangles are $z<0.135$ galaxies, open square are
$z>0.135$ galaxies. Circled and labeled sources are shown in Figure
3. Note the good agreement between the zeropoints of the TF relation
out to $z=0.4$ (left), the clear trend of increasingly negative
TF-offset with bluer color (right), yet the significant scatter about
this correlation of TF residuals with color (right).}
\vskip -0.2in
\end{figure}

We recently acquired high-resolution, WIYN telescope images of a
representative sample of luminous, field spirals at intermediate
redshifts for which we have H$\alpha$ rotation-curves
(e.g., Bershady et al. 1999). We find that significantly over-luminous
galaxies in the $0.1<z<0.4$ $B$-band TF relation tend to appear
optically distorted in the WIYN images -- something we were unable to
determine during the selection process based on lower-resolution
images. Our definition of ``over-luminous'' accounts for the
well-exhibited correlation of TF offsets with color (Figure 2b). Some
of the most extreme outliers exhibiting pronounced optical
distortions are even accompanied by clear signatures of kinematic
asymmetry. A representative sequence is shown in Figure 3.

The physical connection arising here is between bluer colors and
higher luminosity (i.e. enhanced star-formation), which is accompanied
by increased asymmetry at extreme offsets from a fiducial TF
relation. The next step is to establish a statistical trend with
redshift of increased asymmetry and TF offsets above and beyond
changes in stellar M/L. To do so requires further observations. At
intermediate redshifts, higher angular resolution is needed to detect
and quantify the asymmetry in less extreme systems; precision
bi-dimensional spectroscopy is needed to determine reliable
inclinations. Locally, representative kinematic surveys are needed to
understand the link between TF scatter and asymmetry.

\vskip -0.2in

\acknowledgements We gratefully acknowledge our collaborators
M. Haynes, R. Giovanelli, C. Mihos, and D. Koo, C. Conselice,
L. Sparke, J. Gallagher, and E. Wilcots. This research was supported
by NSF/AST-9970780.

\begin{figure}
\plotfiddle{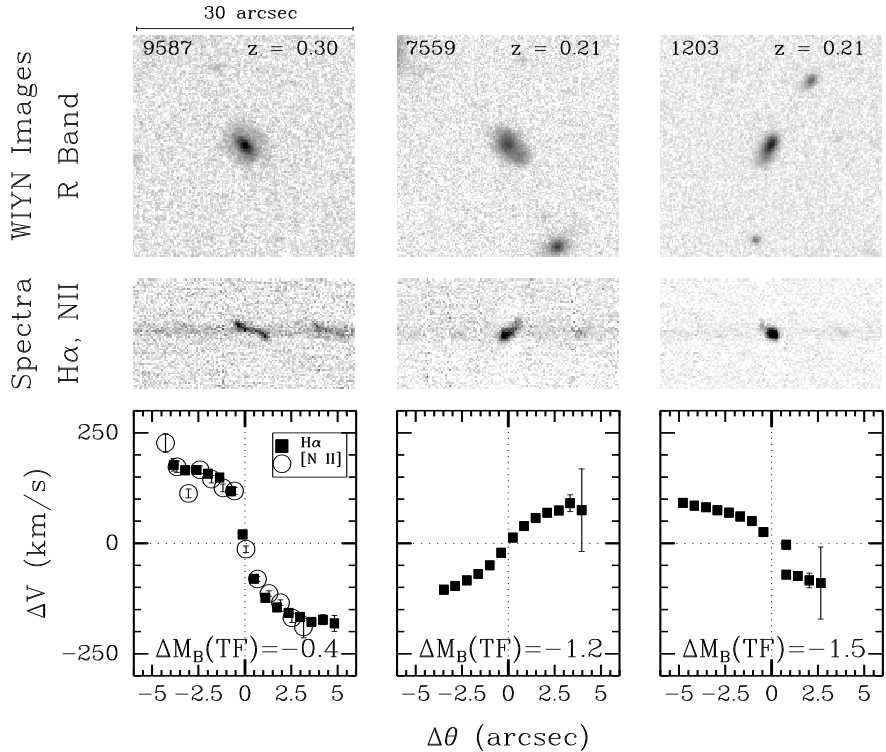}{3in}{-90}{60}{60}{-270}{275}
\vskip 0.5in
\caption{\hsize 5.2in \baselineskip 0.165in A sequence of three
sources in our sample (circled and labeled in Figure 2) with
increasingly blue $B-R$ color, increasingly large photometric and
kinematic asymmetry, and increasingly large (negative) offsets from
the fiducial TF relation (accounting for TF color-dependence).}
\vskip -0.15in
\end{figure}

\vskip -0.35in
\null

\end{document}